\documentclass[12pt,preprint]{aastex}

\usepackage{graphicx}
\usepackage{natbib}
\usepackage{amsmath}
\bibpunct{(}{)}{;}{a}{}{,}

\newcommand{\twofig}[4]{%
  \begin{figure*}%
    \centerline{\resizebox{\hsize}{!}{\includegraphics*{#1} \,%
                                      \includegraphics*{#2}}}%
    \caption{#4}\label{#3}%
  \end{figure*}%
}
\newcommand{\sect}[1]{Sect.~\ref{#1}}
\newcommand{\fig}[1]{Fig.~\ref{#1}}
\newcommand{\eq}[1]{Eq.~\ref{#1}}
\newcommand{\eqs}[1]{Eqs.~\ref{#1}}
\newcommand{\eqq}[1]{Eq.~\ref{#1} et seq.}

\newcommand{\wthreej}[6]{\left( \begin{array}{ccc} #1 & #2 & #3 \\ #4 & #5 & #6 \end{array} \right)}
\newcommand{\gn}{Gaussian}
\newcommand{\ngn}{non-Gaussian}

\newcommand{\ngy}{non-Gaussianity}
\newcommand{\ngs}{non-Gaussianities}
\newcommand{\LL}{{\mathrm{L}}}
\newcommand{\NL}{{\mathrm{NL}}}
\newcommand{\fnl}{$f_{\NL}$}

\begin{document}

\title{Local non-Gaussianity in the Cosmic Microwave Background \\ the
  Bayesian way}
\shorttitle{Local non-Gaussianity in the CMB the Bayesian way}

\author{Franz Elsner}
\affil{Max-Planck-Institut f\"ur Astrophysik,
  Karl-Schwarzschild-Stra\ss e 1, 85748 Garching, Germany}
\email{felsner@mpa-garching.mpg.de}

\and

\author{Benjamin D. Wandelt\altaffilmark{1}}
\affil{UPMC Univ Paris 06, Institut d'Astrophysique de
  Paris, 98 bis, blvd Arago, 75014 Paris, France}
\altaffiltext{1}{Department of Physics, University of Illinois at
  Urbana-Champaign, 1110 W.~Green Street, Urbana, IL 61801, USA}

\begin{abstract}
  We introduce an exact Bayesian approach to search for
  non-Gaussianity of local type in Cosmic Microwave Background (CMB)
  radiation data. Using simulated CMB temperature maps, the newly
  developed technique is compared against the conventional frequentist
  bispectrum estimator. Starting from the joint probability
  distribution, we obtain analytic expressions for the conditional
  probabilities of the primordial perturbations given the data, and
  for the level of \ngy, \fnl, given the data and the
  perturbations. We propose Hamiltonian Monte Carlo sampling as a
  means to derive realizations of the primordial fluctuations from
  which we in turn sample \fnl. Although being computationally
  expensive, this approach allows us to exactly construct the full
  target posterior probability distribution. When compared to the
  frequentist estimator, applying the Bayesian method to \gn\ CMB maps
  provides consistent results. For the analysis of \ngn\ maps,
  however, the error bars on \fnl\ do not show excess variance within
  the Bayesian framework. This finding is of particular relevance in
  the light of upcoming high precision CMB measurements obtained by
  the Planck satellite mission.
\end{abstract}


\section{Introduction}
\label{sec:intro}

Precise measurements of the cosmic microwave background (CMB)
radiation have vastly improved our understanding of cosmology and
played a crucial role in constraining the set of fundamental
cosmological parameters \citep{2003ApJS..148..175S,
  2007ApJS..170..377S, 2009ApJS..180..225H, 2010arXiv1001.4635L}. This
success is based on a tight connection between the temperature
fluctuations we observe today and the physical processes taking place
in the early universe.

Inflation is currently the favored theory predicting the shape of
primordial perturbations \citep{1981PhRvD..23..347G,
  1982PhRvL..48.1220A, 1982PhLB..108..389L, 1982PhLB..117..175S}. In
its simplest form, it is driven by a single scalar field in ground
state with quadratic kinetic term that rolled down a flat potential
slowly. This configuration leads to very small \ngs\ (see
\citealt{2003NuPhB.667..119A, 2003JHEP...05..013M} for a first order,
and \citealt{2010arXiv1003.0481P} for the full second order
calculation). Hence, a clear detection of an excess of primordial
\ngy\ would allow us to rule out the simplest models. Together with
constraints on the scalar spectral index $n_S$ and the search for
primordial gravitational waves, the test for \ngy\ therefore becomes
another important means to probe the physical processes of the early
universe.

In this paper, we focus on non-Gaussianity of local type, where the
amplitude of \ngy\ is measured by a single parameter, \fnl\
\citep{1990PhRvD..42.3936S}. A common strategy for estimating \fnl\ is
to evaluate the bispectrum of the CMB \citep{2002ApJ...566...19K,
  2003ApJS..148..119K, 2007ApJS..170..377S, 2008PhRvL.100r1301Y,
  2009JCAP...09..006S}. This is usually done indirectly via a cubic
combination of filtered CMB maps reconstructing the primordial
perturbations \citep{2005ApJ...634...14K, 2007ApJ...664..680Y,
  2008ApJ...678..578Y}. This approach takes advantage of the specific
signatures produced by primordial \ngy, resulting in a computationally
efficient algorithm. A variant of this estimator has been successfully
applied to the 7-year data release of the Wilkinson Microwave
Anisotropy Probe (WMAP), resulting in $-10 < f_{\NL} < 74$ at $95 \,
\%$ confidence level \citep{2010arXiv1001.4538K}.

The bispectrum estimator used in previous analyses has been shown to
be optimal, i.e.\ it satisfies the Cram\'er-Rao bound
\citep{2005PhRvD..72d3003B}. However, this turns out to be true only
in the limit of vanishing \ngy\ \citep{2007JCAP...03..019C}. For a
significant detection of \fnl, the estimator suffers from excess
variance, a finding that has also been verified numerically
\citep{2007PhRvD..76j5016L}. For the simplified case of a flat sky
approximation, neglected transfer functions and instrumental noise,
\cite{2007JCAP...03..019C} showed that it should be possible to
construct an improved version of the estimator that is equivalent to a
full likelihood analysis up to terms of the order $\mathcal{O}(1/\ln
N_{\mathrm{pix}})$.

Bayesian methods for the analysis of various aspects of CMB data have
been successfully developed in the past, e.g., for an exact power
spectrum determination using Gibbs sampling
\citep[e.g.][]{2004ApJ...609....1J, 2004PhRvD..70h3511W,
  2007ApJ...656..653L, 2009ApJ...697..258J}, to separate foreground
contributions from the CMB anisotropies
\citep[e.g.][]{1998MNRAS.300....1H, 2004MNRAS.351..515B,
  2006NewAR..50..861E, 2008ApJ...672L..87E, 2008ApJ...676...10E,
  2009ApJ...705.1607D}, or to probe for \ngn\ features
\citep[e.g.][]{2001PhRvD..64f3512R, 2009arXiv0911.5399E,
  PhysRevD.80.105005, 2009MNRAS.397..837V}. They offer a natural way
to marginalize over uncertainties e.g.\ attributed to foreground
contamination or instrumental effects. This is of particular
importance for a reliable analysis of weak signals and an advantage
over frequentist methods, where no such procedures exist. Here, we
advance the exact scheme introduced in \citet{2010A&A...513A..59E} to
infer the level of \ngy\ from realistic CMB data within a Bayesian
approach.

We use simulated \gn\ and \ngn\ CMB temperature maps to compare and
contrast the conventional frequentist (bispectrum) estimator with the
exact Bayesian approach. We show that the latter method does not
suffer from excess variance for non-zero \fnl, and can deal with
partial sky coverage and anisotropic noise properties, a feature of
particular importance for local \ngy\ and for any realistic
experiment.

The paper is organized as follows. In \sect{sec:basics}, we briefly
outline the theoretical model used to describe primordial \ngy. We
review the conventional frequentist bispectrum estimator and present
our exact Bayesian approach to infer the amplitude of \ngy\ in
\sect{sec:analysis}. Then, we use simulated maps to compare the
performance of the newly developed technique to the traditional
estimator (\sect{sec:comparison}). We demonstrate the capability of
the Bayesian scheme to deal with realistic CMB experiments in
\sect{sec:wmap}. Finally, we summarize our results in
\sect{sec:summary}.

Throughout the paper, we assume the WMAP5+BAO+SNALL cosmological
parameters \citep{2009ApJS..180..330K}: $\Omega_{\Lambda} = 0.721$,
$\Omega_{c} \, h^2 = 0.1143$, $\Omega_{b} \, h^2 = 0.02256$,
$\Delta^2_{\mathcal{R}}(0.002 \ Mpc^{-1}) = 2.457 \cdot 10^{-9}$, $h =
0.701$, $n_{s} = 0.96$, and $\tau = 0.084$.

\section{Model of non--Gaussianity}
\label{sec:basics}

The multipole coefficients $a_{\ell m}$ of the CMB temperature
anisotropies are related to the primordial fluctuations,
\begin{align}
  \label{eq:phi2alm}
  a_{\ell m}&=\frac{2}{\pi}\int k^2 dk \, r^2 dr \, \int d\Omega \,
  \Phi(\hat{n},r) Y_{\ell m}^{*} \; g_{\ell}(k) \, j_\ell(kr) +
  n_{\ell m} \nonumber \\ &\equiv M \Phi_{\ell m} + n_{\ell m} \, ,
\end{align}
where $\Phi_{\ell m}$ is the spherical harmonic transform of the
primordial adiabatic perturbations at comoving distance $r$,
$g_{\ell}$ the transfer function in momentum space, and $j_\ell$ the
spherical Bessel function of order $\ell$. Additive noise is taken
into account by $n_{\ell m}$, for a compact notation we will use the
operator $M$ as a shorthand for the radial integral in what
follows. Traces of \ngy\ in the primordial fluctuations will be
transferred to the multipole moments $a_{\ell m}$ according to
\eq{eq:phi2alm}, potentially making them accessible to CMB
experiments.

We focus on non-Gaussianity of local type, which is realized to very
good approximation in multi-field inflationary models as described by
the curvaton model \citep{2001PhLB..522..215M, 2003PhRvD..67b3503L},
or in ekpyrotic/cyclic universe models \citep{2001PhRvD..64l3522K,
  2002NuPhB.626..395E, 2002PhRvD..65l6003S}. Here, we can parametrize
the \ngy\ of $\Phi$ via a quadratic dependency on a \gn\ auxiliary
field $\Phi_{\LL}$, that is local in real space, of the form
\citep{1990PhRvD..42.3936S, 1994ApJ...430..447G}
\begin{equation}
\label{eq:fnldef}
\Phi(r)=\Phi_{\LL}(r) + f_{\NL}\lbrack \Phi^2_{\LL}(r)-\langle
\Phi^2_{\LL}(r) \rangle \rbrack + {\cal O}(\Phi^3_{\LL}) \, ,
\end{equation}
where \fnl\ is a dimensionless measure of the amplitude of \ngy\ and
we truncate the expansion at third order in $\Phi_{\LL}$.

The Bayesian method presented in the following section takes advantage
of the simple form of \eq{eq:fnldef}, which links the properties of
the primordial perturbations $\Phi$ to that of a \gn\ random field
$\Phi_{\LL}$. As a result, it cannot easily be generalized to the
analysis of other types of \ngy, where no such relation exists. Though
this poses an important limitation of the method, improved statistical
means for the search for \ngy\ of local type are of particular
relevance as the conventional bispectrum estimator is known to suffer
from large excess variance here. Finally, we explicitly stress the
interesting possibility to include the cubic term $\Phi_{\LL}^3$ in
the perturbational expansion (\eq{eq:fnldef}) to obtain simultaneously
constraints to the next order \ngy\ parameter, commonly referred to as
$g_{\NL}$.

\section{Analysis techniques}
\label{sec:analysis}

\subsection{Frequentist estimator}

In the following, we briefly review the fast estimator as proposed by
\citet{2005ApJ...634...14K}. This estimator is optimal for uniform
observation of the full sky. More general least-square cubic
estimators have been found for data with partial sky coverage and
anisotropic noise (\citealt{2006JCAP...05..004C}, see also the review
of, e.g., \citealt{2010arXiv1006.0275Y}).

To estimate the \ngy\ of a CMB temperature map, one constructs the
statistic $\mathcal{S}_{prim}$ out of a cubic combination of the data,
\begin{equation}
  \mathcal{S}_{prim} = \int dr \, r^2 \ \int d^2\hat{n} \ A(r, \hat{n}) \
  B^2(r, \hat{n}) \, .
\end{equation}
The spatial integral runs over two filtered maps,
\begin{align}
 \label{eq:A_map}
 A(r, \hat{n}) &= \sum_{\ell, m}
 \mathcal{C}^{-1}_{\ell} \ \alpha_{\ell}(r) \ a_{\ell m} \
 Y_{\ell m}(\hat{n}) \, ,\\
 \label{eq:B_map}
 B(r, \hat{n}) &= \sum_{\ell, m}
 \mathcal{C}^{-1}_{\ell} \ \beta_{\ell}(r) \ a_{\ell m} \
 Y_{\ell m}(\hat{n}) \, ,
\end{align}
that are constructed using the auxiliary functions
\begin{align}
  \alpha_\ell(r) &= \frac{2}{\pi} \int dk \, k^2 \ g_\ell(k) \ j_\ell(kr) \, ,\\
 \beta_{\ell}(r) &= \frac{2}{\pi} \int dk \, k^2 \mathcal{P}(k) \
 g_{\ell}(k) \ j_{\ell}(k r) \, ,
\end{align}
and the inverse of the CMB plus noise power spectrum,
$\mathcal{C}^{-1}_{\ell}$. The power spectrum of the primordial
perturbations is denoted by $\mathcal{P}(k)$. Now, we can calculate
the estimated value of \fnl\ from the statistics $\mathcal{S}_{prim}$
by applying a suitable normalization,
\begin{equation}
  \label{eq:fast_est}
  \hat{f}_{\NL} = \left[ \sum_{\ell_1 \le \ell_2 \le \ell_3} \ \frac{1}{\Delta_{\ell_1 \ell_2 \ell_3}}
    \ (B^{prim})^2_{\ell_1 \ell_2 \ell_3} \ \mathcal{C}^{-1}_{\ell_1} \ \mathcal{C}^{-1}_{\ell_2} \
    \mathcal{C}^{-1}_{\ell_3} \right]^{-1} \! \cdot \ \mathcal{S}_{prim} \, ,
\end{equation}
where $\Delta_{\ell_1 \ell_2 \ell_3} = 6$, when $\ell_1 = \ell_2 =
\ell_3$, 2, when $\ell_1 = \ell_2 \neq \ell_3$ or $\ell_1 \ne \ell_2 =
\ell_3$, and 1 otherwise. The theoretical bispectrum for $f_{\NL} = 1,
\ B^{prim}_{\ell_1 \ell_2 \ell_3}$, is given by
\begin{equation}
  B^{prim}_{\ell_1 \ell_2 \ell_3}= 2 \, I_{\ell_1 \ell_2 \ell_3}
  \int dr \, r^2 \ \lbrack \beta_{\ell_1}(r) \beta_{\ell_2}(r)
  \alpha_{\ell_3}(r) + \beta_{\ell_3}(r) \beta_{\ell_1}(r)
  \alpha_{\ell_2}(r) + \beta_{\ell_2}(r) \beta_{\ell_3}(r)
  \alpha_{\ell_1}(r) \rbrack \, ,
\end{equation}
where a combinatorial prefactor is defined as
\begin{equation}
  I_{\ell_1 \ell_2 \ell_3} = \sqrt{\frac{(2 \ell_1 + 1)(2 \ell_2 +
      1)(2 \ell_3 + 1)}{4 \pi}} \,
  \wthreej{\ell_1}{\ell_2}{\ell_3}{0}{0}{0} \, .
\end{equation}

Recently, the Bayesian counterpart of the fast estimator has been
developed within the framework of information field theory by
expanding the logarithm of the posterior probability to second order
in \fnl\ \citep{PhysRevD.80.105005}. Here, the equivalent of the
normalization factor in \eq{eq:fast_est} becomes data dependent,
accounting for the fact that the ability to constrain \fnl\ varies
from data set to data set. We will go beyond this level of accuracy
and present an exact Bayesian scheme in the next section.

\subsection{Exact Bayesian inference}

We now introduce a Bayesian method that, in contrast to the bispectrum
estimator, includes information from \emph{all} correlation
orders. Our aim is to construct the posterior distribution of the
amplitude of \ngs\ given the data, $P(f_{\NL} | d)$. To this end, we
subsume the remaining set of cosmological parameters to a vector
$\theta$ and rewrite the joint distribution as
\begin{equation}
  P(d, \Phi_{\LL}, f_{\NL}, \theta) = P(d | \Phi_{\LL}, f_{\NL}, \theta) \
  P(\Phi_{\LL} | \theta) \ P(f_{\NL}) \ P(\theta) \, .
\end{equation}
Substituting the noise vector in terms of data and signal, we can use
\eqq{eq:phi2alm} to express the probability for data $d$ given
$\Phi_{\LL}$, \fnl, and $\theta$ up to an overall prefactor
\begin{equation}
  P(d | \Phi_{\LL}, f_{\NL}, \theta) \propto \mathrm{e} \, ^{ {-1/2 \,
      \lbrack d - M ( \Phi_{\LL} + f_{\NL} (\Phi^2_{\LL} - \langle
      \Phi^2_{\LL} \rangle)) \rbrack^{\dagger} N^{-1} \lbrack d - M (
      \Phi_{\LL} + f_{\NL} (\Phi^2_{\LL} - \langle \Phi^2_{\LL}
      \rangle) ) \rbrack } }\, , 
\end{equation}
where we introduced the noise covariance matrix $N$. The prior
probability $P(\Phi_{\LL} | \theta)$ can be expressed as multivariate
Gaussian distribution by construction, thus, we eventually obtain
\begin{multline}
  \label{eq:jointdistr}
  P(d, \Phi_{\LL}, f_{\NL}, \theta) \propto \exp \left\{ -1/2 \, \left[d
      - M (\Phi_{\LL} + f_{\NL} (\Phi^2_{\LL} - \langle \Phi^2_{\LL}
      \rangle))\right]^{\dagger} N^{-1} \right. \\
  \left. \times \left[d - M ( \Phi_{\LL} + f_{\NL} (\Phi^2_{\LL} - \langle
    \Phi^2_{\LL} \rangle))\right] -1/2 \, \Phi_{\LL}^{\dagger}
  P^{-1}_{\Phi} \Phi_{\LL} - f_{\NL}^2/2 \sigma_{f_{\NL}}^2
\right\}
\end{multline}
as an exact expression for the joint distribution up to a
normalization factor, assuming a \gn\ prior for \fnl\ with zero mean
and variance $\sigma_{f_{\NL}}^2$, and a flat prior for the
cosmological parameters. The covariance matrix $P_{\Phi}$ is
constrained by the primordial power spectrum predicted by inflation,
$\mathcal{P}(k)$, and given by \citep{2003ApJ...597...57L}
\begin{equation}
  \label{eq:cov_phi}
  \left\langle \Phi_{\LL \ \ell_1 m_1}(r_1) \, \Phi^{*}_{\LL \ \ell_2 m_2}(r_2)
  \right\rangle = \frac{2}{\pi} \ \delta_{\ell_2}^{\ell_1}\ \delta_{m_2}^{m_1}
  \ \int dk \ k^2 \mathcal{P}(k) \ j_{\ell_1}(kr_1) \ j_{\ell_2}(kr_2) \, .
\end{equation}

To evaluate the joint distribution (\eq{eq:jointdistr}) directly would
require to perform a numerical integration over a high dimensional
parameter space. For realistic data sets this turns out to be
impossible computationally. We pursue a different approach
here. First, we note that the exponent in \eq{eq:jointdistr} is
quadratic in \fnl\ and hence the conditional density $P(f_{\NL} | d,
\Phi_{\LL}, \theta)$ is Gaussian with mean and variance
\begin{align}
  \label{eq:fnl_sample}
  \langle f_{\NL} \rangle &= \langle (f_{\NL} - \langle f_{\NL}
  \rangle)^2 \rangle (\Phi^2_{\LL} - \langle \Phi^2_{\LL}
  \rangle)^{\dagger} M^{\dagger} N^{-1} (d - M \Phi_{\LL})  \nonumber \\
  \langle (f_{\NL} - \langle f_{\NL} \rangle)^2 \rangle &= \left[
    (\Phi^2_{\LL} - \langle \Phi^2_{\LL} \rangle)^{\dagger}
    M^{\dagger} N^{-1} M (\Phi^2_{\LL} - \langle \Phi^2_{\LL} \rangle)
    + 1/\sigma_{f_{\NL}}^2 \right]^{-1} \, .
\end{align}
Thus, for any realization of $\Phi_{\LL}$, \eqs{eq:fnl_sample} permit
us to calculate the distribution of \fnl\ given the data. Similarly,
we can calculate the conditional probability $P(\Phi_{\LL} | d,
\theta)$ by analytically marginalizing \eq{eq:jointdistr} over \fnl,
\begin{align}
  \label{eq:phi_sample}
  P(\Phi_{\LL} | d, \theta) &= \int d f_{\NL} \, P(\Phi_{\LL},f_{\NL}
  | d, \theta) \nonumber \\ &\propto [ \sigma_{f_{\NL}}^2(\Phi^2_{\LL}
    - \langle \Phi^2_{\LL} \rangle)^{\dagger} M^{\dagger} N^{-1} M
    (\Phi^2_{\LL} - \langle \Phi^2_{\LL} \rangle) + 1 ]^{-1/2}
  \nonumber \\ & \quad \times \mathrm{e} \, ^{-1/2 \, (d - M
    \Phi_{\LL})^{\dagger} \left[ N^{-1} - \frac{\sigma_{f_{\NL}}^2
        N^{-1} M (\Phi^2_{\LL} - \langle \Phi^2_{\LL} \rangle)
        (\Phi^2_{\LL} - \langle \Phi^2_{\LL} \rangle)^{\dagger}
        M^{\dagger} N^{-1} }{\sigma_{f_{\NL}}^2(\Phi^2_{\LL} - \langle
        \Phi^2_{\LL} \rangle)^{\dagger} M^{\dagger} N^{-1} M
        (\Phi^2_{\LL} - \langle \Phi^2_{\LL} \rangle) + 1} \right] }
    \nonumber \\
    & \qquad ^{ \times (d - M \Phi_{\LL}) -1/2 \, \Phi_{\LL}^{\dagger} P^{-1}_{\Phi}
    \Phi_{\LL} } .
\end{align}
Now we can outline our approach to infer the level of \ngy\ from CMB
data iteratively. First, for given data $d$, we draw $\Phi_{\LL}$ from
the distribution \eq{eq:phi_sample}. Then, \fnl\ can be sampled
according to \eqs{eq:fnl_sample} using the value of $\Phi_{\LL}$
derived in the preceding step. If the sampling scheme is iterated for
a sufficient amount of cycles, the derived set of \fnl\ values
resembles an unbiased representation of the posterior distribution
$P(f_{\NL} | d, \theta)$.

Unfortunately, there exists no known way to draw uncorrelated samples
of $\Phi_{\LL}$ from its \ngn\ distribution function directly. Here,
we propose Hamiltonian Monte Carlo (HMC) sampling to obtain correlated
realizations of the primordial perturbations. Contrary to conventional
Metropolis-Hastings algorithms, it avoids random walk behavior in
order to increase the acceptance rate of the newly proposed
sample. This is a mandatory requirement to explore successfully
high-dimensional parameter spaces as found here. For HMC sampling, the
variable is regarded as the spatial coordinate of a particle moving in
a potential well described by the probability distribution function to
evaluate \citep{Duane1987}. A generalized mass matrix $W$ and momentum
variables $p$ are assigned to the system to define its Hamiltonian
\begin{equation}
  H = 1/2 \, p^{\dagger}W^{-1}p - \log[ P(\Phi_{\LL} | d, \theta) ] \, ,
\end{equation}
where the potential is related to the posterior distribution as
defined in \eq{eq:phi_sample}. The system is evolved deterministically
from a starting point according to the Hamilton's equations of motion
\begin{align}
\frac{{\mathrm{d}}\Phi_{\LL}}{\mathrm{dt}} &= \frac{\partial H}{\partial
  p} \, , \nonumber \\ \frac{{\mathrm{d}}p}{\mathrm{dt}} &=
-\frac{\partial H}{\partial {\Phi_{\LL}}} = \frac{\partial
  \log[P(\Phi_{\LL} | d, \theta)]}{\partial {\Phi_{\LL}}} \, ,
\end{align}
which are integrated by means of the second order leapfrog scheme with
step size $\delta t$,
\begin{align}
  p(t + \frac{\delta t}{2}) &= p(t) + \frac{\delta t}{2} \,
  \frac{\partial \log[P(\Phi_{\LL} | d, \theta)]}{\partial \Phi_{\LL}}
  \bigg|_{\Phi_{\LL}(t)} \nonumber \\ \Phi_{\LL}(t + \delta t) &=
  \Phi_{\LL}(t) +
  \delta t \, W^{-1} p(t + \frac{\delta t}{2}) \nonumber \\
  p(t + \delta t) &= p(t + \frac{\delta t}{2}) + \frac{\delta
    t}{2} \, \frac{\partial \log[P(\Phi_{\LL} | d, \theta)]}{\partial
    {\Phi_{\LL}}}\bigg|_{\Phi_{\LL}(t + \delta t)} \, .
\end{align}
The equation of motion for $\Phi_{\LL}$ can easily be solved, as it
only depends on the momentum variable. To integrate the evolution
equation for $p$, we derive
\begin{multline}
  \label{eq:hmc_grad}
  \frac{\partial \log[P(\Phi_{\LL} | d, \theta)]}{\partial
    {\Phi_{\LL}}} \approx M^{\dagger} \left[ N^{-1} -
    \frac{\sigma_{f_{\NL}}^2 N^{-1} M (\Phi^2_{\LL} - \langle
      \Phi^2_{\LL} \rangle) (\Phi^2_{\LL} - \langle \Phi^2_{\LL}
      \rangle)^{\dagger} M^{\dagger} N^{-1}
    }{\sigma_{f_{\NL}}^2(\Phi^2_{\LL} - \langle \Phi^2_{\LL}
      \rangle)^{\dagger} M^{\dagger} N^{-1} M (\Phi^2_{\LL} - \langle
      \Phi^2_{\LL} \rangle) + 1} \right] \\
  \times (d - M \Phi_{\LL}) - P_\Phi^{-1} \Phi_{\LL}
\end{multline}
as an approximate expression neglecting higher order terms in
$\Phi_{\LL}$. The final point of the trajectory is accepted with
probability $a = \min(1, \exp[-\Delta H ])$, where $\Delta H$ is the
difference in energy between the end- and starting point. As the
energy is conserved in a system with time-independent Hamiltonian, the
acceptance rate in case of an exact integration of the equations of
motion would be unity, irrespecticive of the complexity of the
problem.  Introducing the accept/reject step restores exactness also
in realistic applications as it eliminates the error originating from
approximating the gradient in \eq{eq:hmc_grad} and from the numerical
integration scheme. In general, only accurate integrations where
$\Delta H$ is close to zero result in high acceptance rates. This can
usually be archived by choosing small time steps or an accurate
numerical integration scheme. However, as the time integration
requires the calculation of spherical harmonic transforms with
inherently limited precision, higher order methods turn out to be
unrewarding. Furthermore, the efficiency of a HMC sampler is sensitive
to the choice of the mass matrix $W$. In agreement with
\citet{2008MNRAS.389.1284T}, we found best performance when choosing
$W$ as inverse of the posterior covariance matrix of the primordial
perturbations, which we derive from the Wiener filter equation for
purely \gn\ perturbations to good approximation,
\begin{align}
  P(d, \Phi^{G}, \theta) &\propto \exp \left\{ -1/2 \, \left[d - M \Phi^{G}
    \right]^{\dagger} N^{-1} \left[d - M \Phi^{G} \right] -1/2 \,
    \Phi^{G \, \dagger} P^{-1}_{\Phi} \Phi^{G} \right\} \, ,
\end{align}
with mean and variance of the distribution $P(\Phi^{G} | d, \theta)$
\begin{align}
  \label{eq:wf}
  \langle \Phi^{G} \rangle &= \langle (\Phi^{G} - \langle \Phi^{G}
  \rangle)^2 \rangle M^{\dagger} N^{-1} d \nonumber  \\
  \langle (\Phi^{G} - \langle \Phi^{G} \rangle)^2 \rangle &= \left[
    M^{\dagger} N^{-1} M - P^{-1}_{\Phi} \right]^{-1} \, , \text{ hence}\\
  W &= M^{\dagger} N^{-1} M - P^{-1}_{\Phi} \, .
\end{align}
For the calculation of the mass matrix $W$ in the presence of
anisotropic noise or partial sky coverage, we still adopt a simple
power spectrum as approximation for $N^{-1}$ in spherical harmonic
space at the cost of a reduced sampling efficiency.

We initialize the algorithm by performing one draw of the primordial
perturbations from the \gn\ posterior $P(\Phi^{G} | d, \theta)$
(\eqs{eq:wf}).

\section{Scheme comparison}
\label{sec:comparison}

We use simulated CMB temperature maps obtained with the algorithm
described in \citet{2009ApJS..184..264E} to compare the newly
developed Bayesian scheme to the conventional frequentist approach. We
chose a \gn\ ($f_{\NL} = 0$) and a \ngn\ ($f_{\NL} = 100$) CMB
realization at a HEALPix resolution of $n_{side} = 256$ and
$\ell_{max} = 512$, superimposed by isotropic noise with a constant
power spectrum amplitude of $\mathcal{C}^{noise}_{\ell} = 10^{-7}
\mathrm{mK}^2$. We show the \ngn\ temperature map besides the input
signal and noise power spectra in \fig{fig:sim_map}.

Performing the analysis within the frequentist framework, we derive
$\hat{f}_{\NL} = 4$ for the \gn\ and $\hat{f}_{\NL} = 97$ for the
\ngn\ simulation. To obtain an estimate of the attributed error, we
conducted 1000 Monte Carlo simulations with the input parameters as
quoted above. For the \gn\ realization, we find a standard deviation
of $\sigma_{f_{\NL}}^{MC} = 15$, in perfect agreement with the value
predicted form a fisher information matrix forecast. For the \ngn\
simulation, however, the derived error $\sigma_{f_{\NL}}^{MC} = 20$ is
already considerably larger than in the \gn\ case---the sub-optimality
of the bispectrum estimator at non-zero \fnl\ becomes manifest.

In the Bayesian analysis, we construct the full posterior distribution
out of the samples drawn from it. We chose a \gn\ prior for \fnl\ with
zero mean and a very large width of $\sigma_{f_{\NL}}^{prior} = 500$
in order to not introduce any bias to the results. For an efficient
sampling process, we tuned the time step size $\delta t$ of the HMC
algorithm to realize a mean acceptance rate of about $40 \, \%$. To
reduce the overall wall clock time needed for the analysis of one CMB
map, we ran 32 chains in parallel and eventually combine all the
samples. For reliable results, it is imperative to quantitatively
assess the convergence of the Monte Carlo process. Here, we apply the
statistics of \citet{199211} to the obtained samples. It compares the
variance among different chains with the variance within a chain and
returns a number in the range of $0 \le R < \infty$ which reflects the
quality of the convergence of the chains with a given length. In
general, a value close to $R = 1$ reflects good convergence. As this
value refers to the convergence of a single chain, we in fact obtain a
significantly better result after a combination of all of the 32
independent chains we generated.

For the \gn\ simulation, we run chains with a length of $25 \, 000$
samples each, discarding the first 5000 samples during burn-in. With
these parameters, we find excellent convergence as confirmed by the
Gelman-Rubin statistics, $R = 1.04$. The final result along with a
comparison to the frequentist scheme is shown in
\fig{fig:g_analysis}. In the Bayesian analysis, we find a mean value
of $\langle f_{\NL} \rangle = 3$ and a width of the distribution
$\sigma_{f_{\NL}} = 15$. As the bispectrum estimator is known to be
optimal in the limit of vanishing \ngy, the two different approaches
lead to consistent results.

To repeat the analysis of the \ngn\ map, we again generated 32
independent chains with a length of $40 \, 000$ samples each. After
dropping the first $10 \, 000$ elements to account for the period of
burn-in, we estimated the convergence of the individual chains by
means of the Gelman-Rubin statistics and find $R = 1.4$. The inferred
mean of $\langle f_{\NL} \rangle = 99$ at an 1-$\sigma$ error of
$\sigma_{f_{\NL}} = 15$ is in good agreement with the input value of
the simulation. We directly compare the Bayesian to the frequentist
result in \fig{fig:ng_analysis}, where we now find an important
difference in the outcomes. Whereas for a significant detection of
\ngy\ the frequentist estimator suffers from excess variance, the
Bayesian scheme still provides the same error bars as for the
\gn\ simulation. This increase in variance has been found to be an
intrinsic property of the conventional bispectrum estimator applied to
the detection of local \ngy. \citet{2007JCAP...03..019C} show the
existence of an improved cubic estimator which better approximates the
maximum likelihood estimator even for non-vanishing values of
\fnl. While this estimator has not yet been constructed for realistic
data sets, the Bayesian analysis we present here yields as a
by-product the maximum a posteriori estimator which becomes the
maximum likelihood estimator in the limit of large prior variance for
\fnl. In addition, the Bayesian analysis produces the full posterior
distribution using all the information about \fnl\ contained in the
data. As we demonstrate in this paper, the variance of the posterior
distribution does not change in the case of non-zero \fnl, but its
shape does.

We note that the computational cost for the Bayesian analysis with the
exact marginalization of the high-dimensional $\Phi$ parameter space
is quite demanding. With the setup as described here, the runtime for
the \gn\ and the \ngn\ simulation amounts to about $80 \, 000$ CPUh
and $150 \, 000$ CPUh, respectively. It is dominated by spherical
harmonic transforms that show a scaling behavior of
$\mathcal{O}(N_{\mathrm{pix}}^{3/2})$, where $N_{\mathrm{pix}}$ are
the number of pixels in the data map. Though computationally
expensive, the algorithm in its present implementation enables the
analysis of WMAP data with an only moderately higher resolution than
that of the simulations considered here. The reason for the
inefficiency of the algorithm lies in the large correlation length of
the \fnl\ sampling chains. We illustrate this fact in
\fig{fig:hmc_performance}, where we display three out of the 32 chains
of the \ngn\ simulation. In addition, we show the autocorrelation
function of a chain as defined via
\begin{equation}
  \xi (\Delta N) = \frac{1}{N} \sum^{N}_{i} \frac{(f_{\NL}^i - \mu)
    (f_{\NL}^{i + \Delta N} - \mu)}{\sigma^2} \, ,
\end{equation}
where N is the length of the \fnl\ chains with mean $\mu$ and variance
$\sigma^2$.

It is interesting to note that the derived values of \fnl\ and their
error bars will in general not agree exactly between the two
approaches, even for a \gn\ data set. The frequentist estimator is
unbiased with respect to all possible realizations of signal and
noise. The error bars, calculated via Monte Carlo simulations, are the
same for all data sets with identical input parameters by
definition. The Bayesian approach, on the other hand, returns the
entire information contained about the local model in the particular
realization subject to the analysis. Thus, the uncertainty in the
parameter is computed from the data itself and will vary from data set
to data set, as cosmic variance or accidental alignments between
signal and noise may impact the ability to constrain the level of
\ngy. Furthermore, the Bayesian method constructs the full posterior
probability function instead of simply providing an estimate of the
error under the implicit assumption of a \gn\ distribution.

\twofig{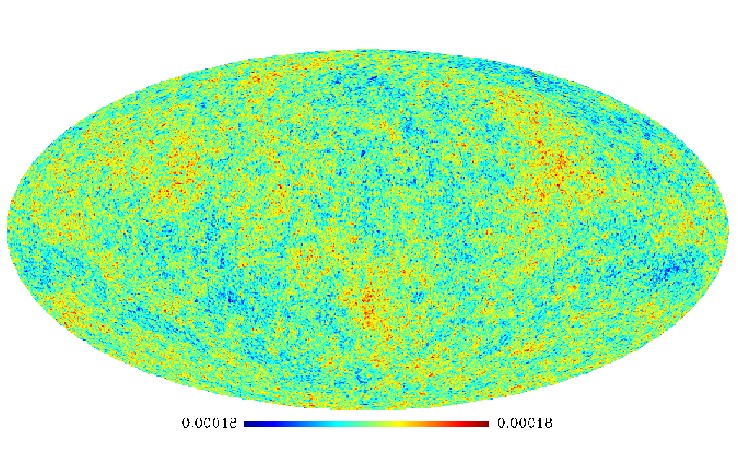}{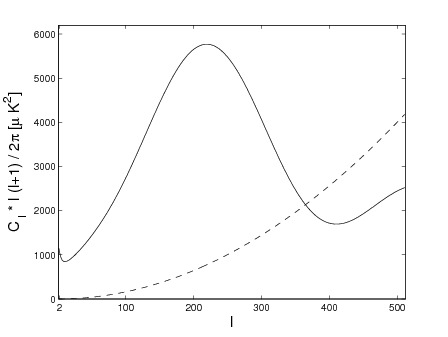}{fig:sim_map}%
{Properties of the maps analyzed. \emph{Left panel:} Our non-Gaussian
  CMB signal simulation in dimensionless units. \emph{Right panel:}
  The input signal (solid line) and noise (dashed line) power spectra.}

\twofig{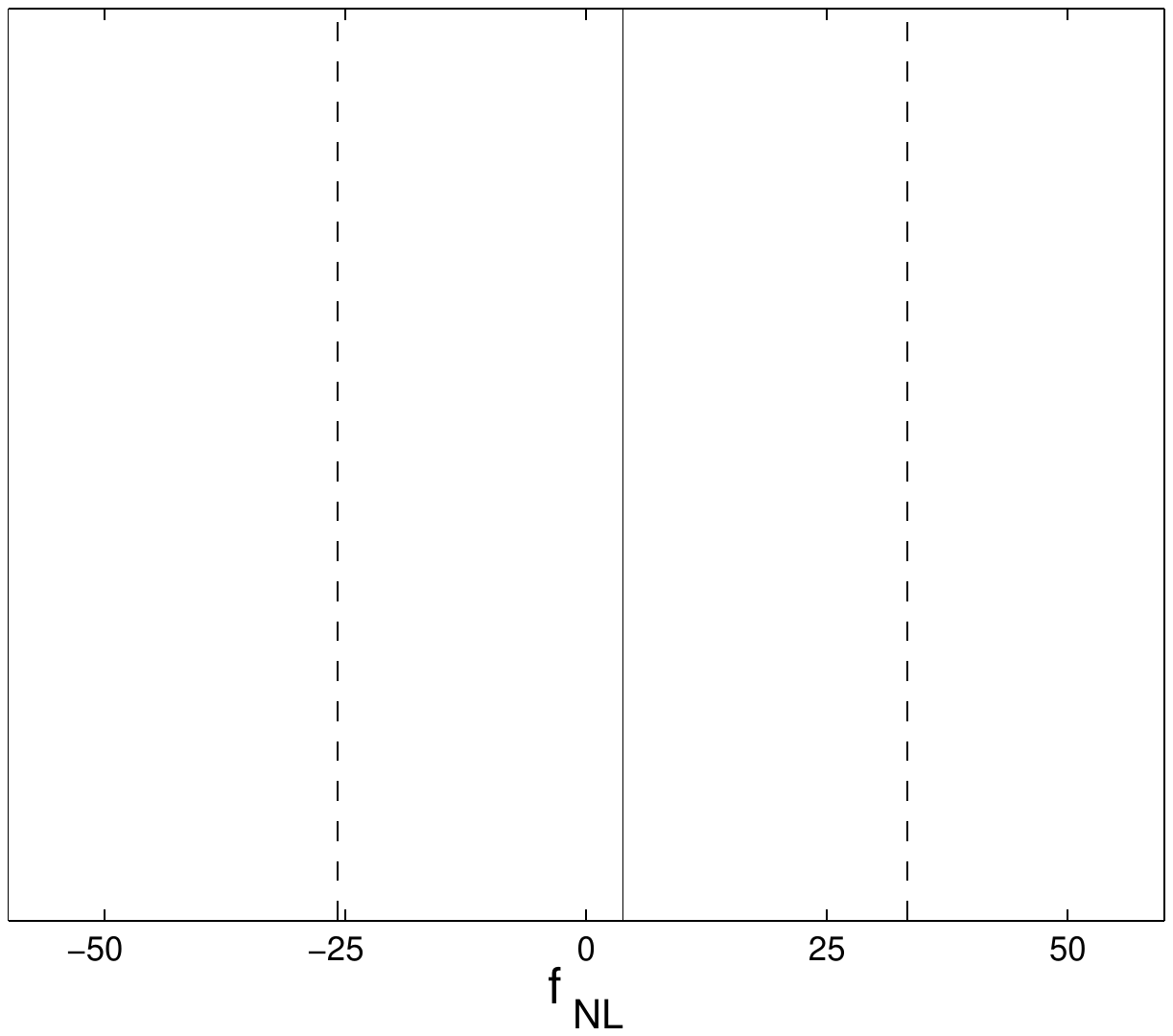}{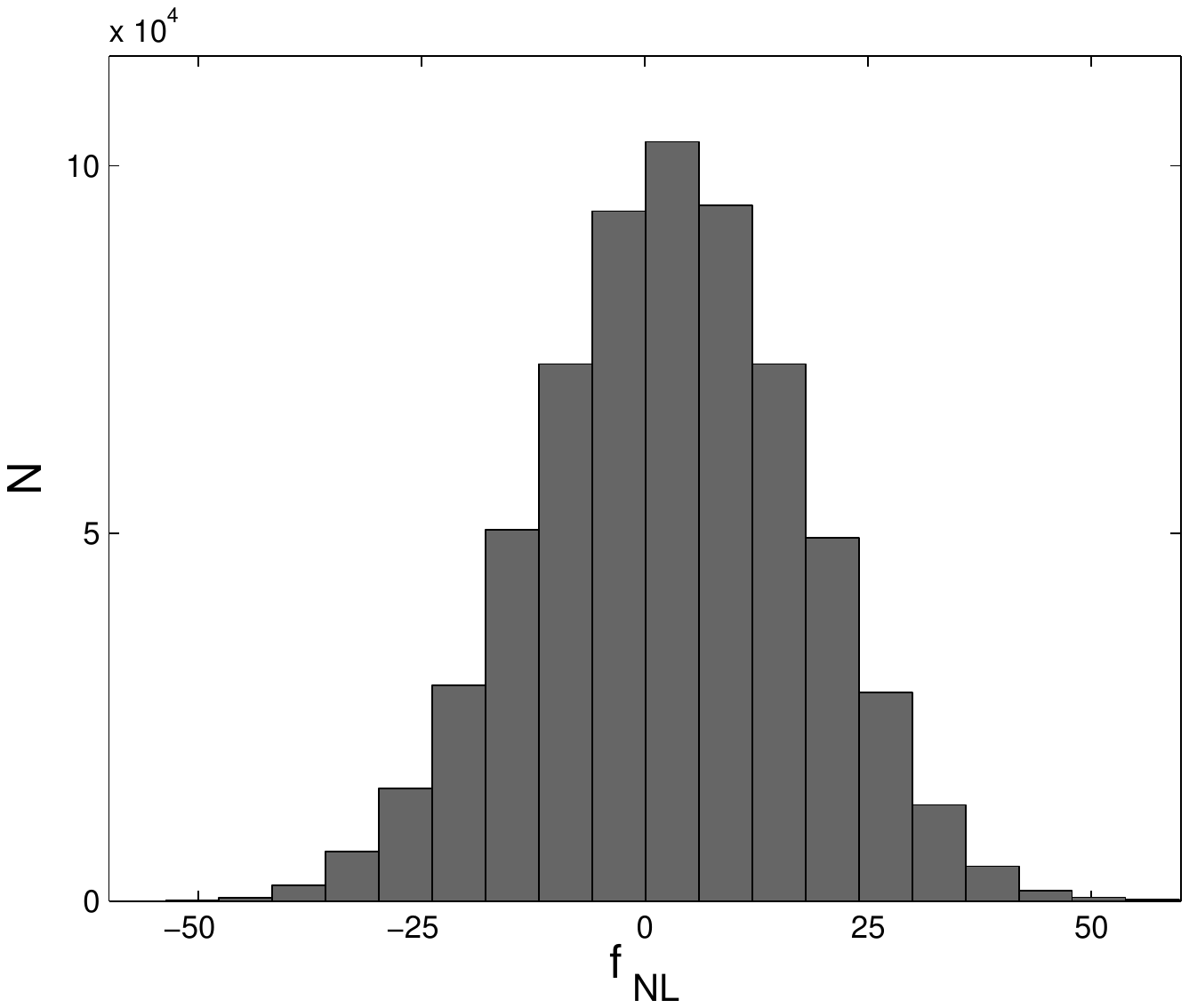}{fig:g_analysis}%
{Analysis of the \gn\ simulation ($f_{\NL} = 0$). \emph{Left panel:}
  We show the analysis of the \gn\ CMB map by means of the frequentist
  estimator. Plotted are the recovered value $\hat{f}_{\NL} = 4$
  (\emph{solid line}) and the $2-\sigma$ error (\emph{dashed lines})
  as derived from Monte Carlo simulations, $\sigma_{f_{\NL}}^{MC} =
  15$. \emph{Right panel:} The analysis of the same data set within a
  Bayesian framework constructs the full posterior distribution
  $P(f_{\NL} | d, \theta)$. We obtain a mean value of $\langle f_{\NL}
  \rangle = 3$ and a standard deviation of $\sigma_{f_{\NL}} = 15$.}

\twofig{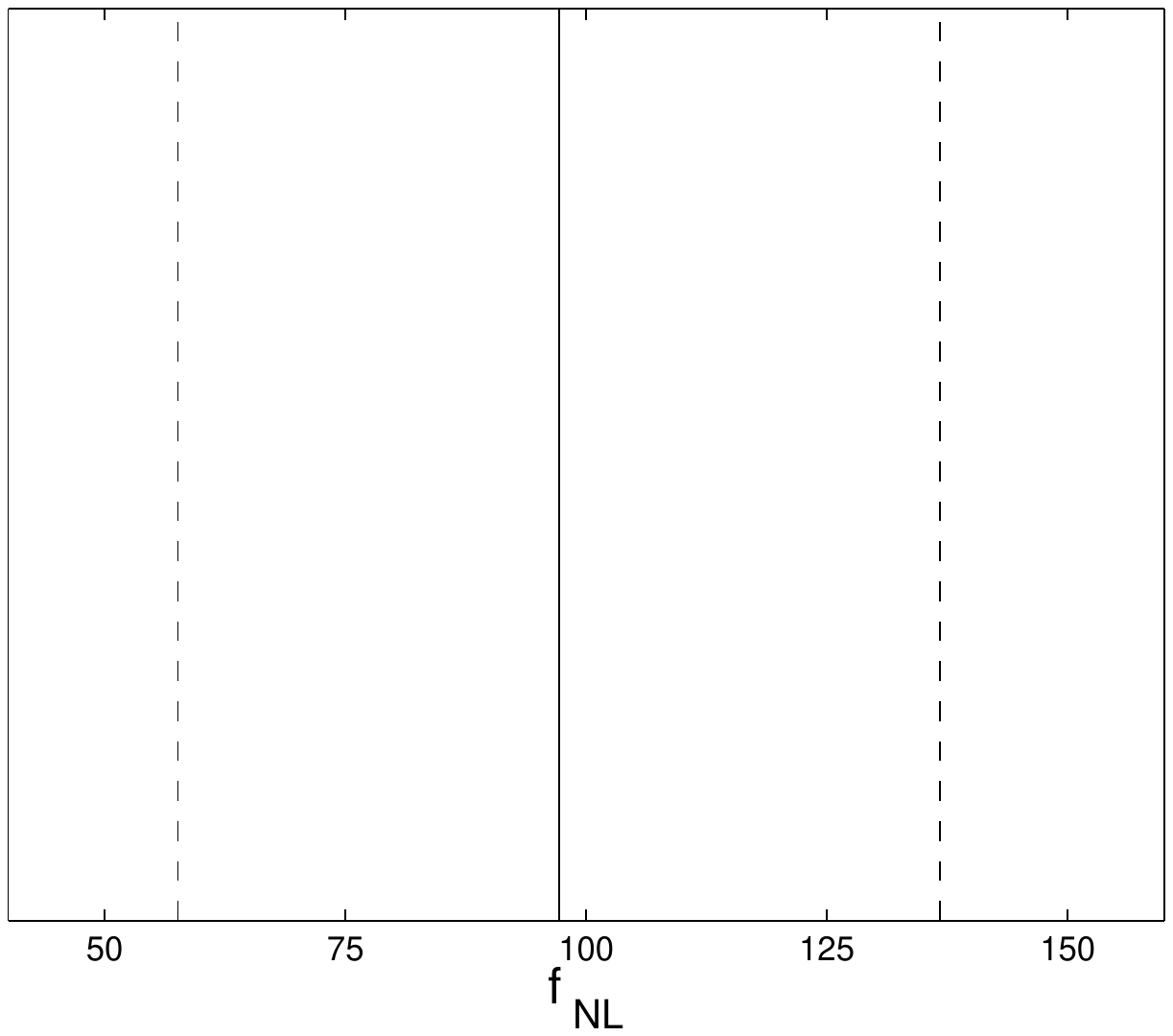}{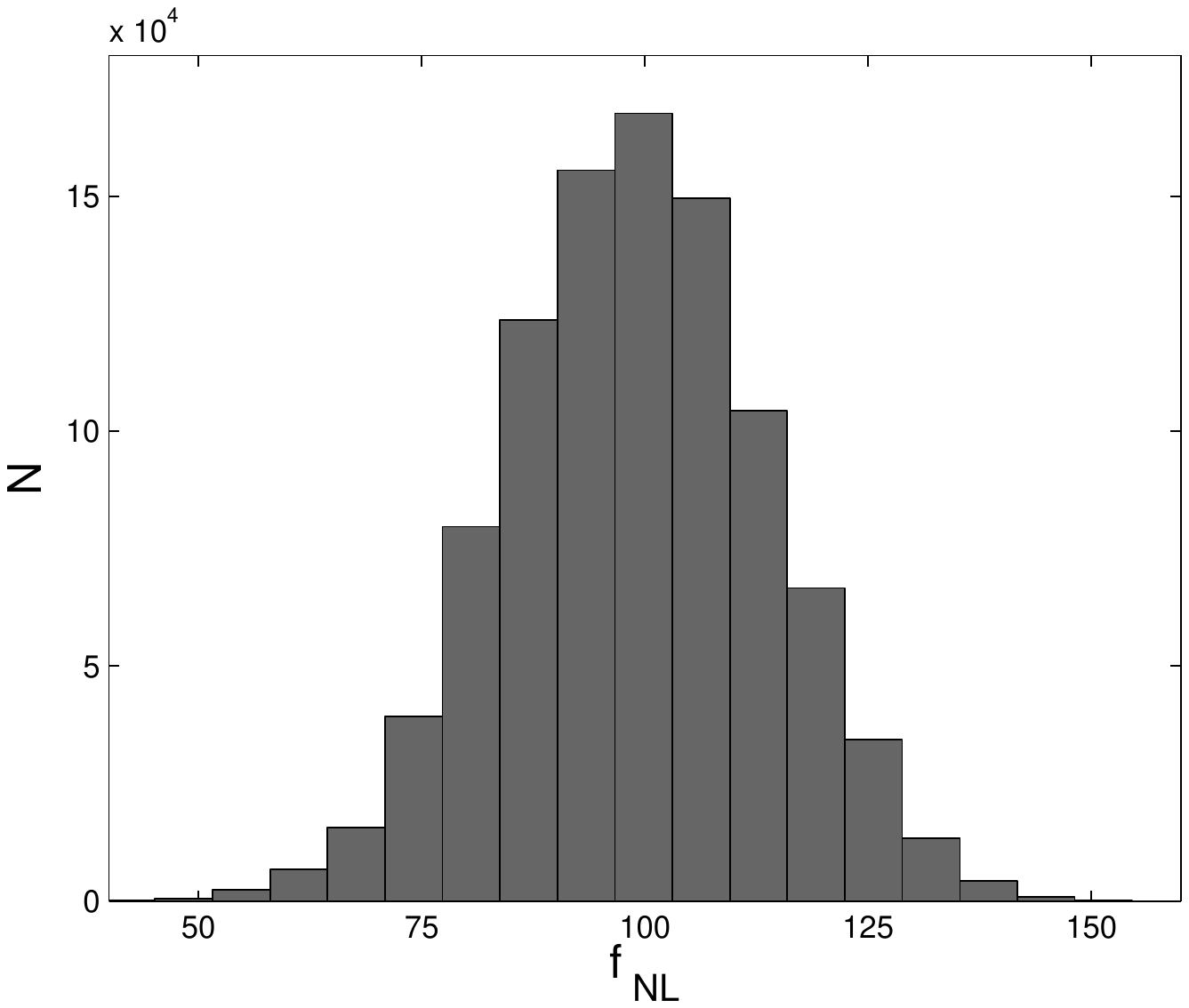}{fig:ng_analysis}%
{Same as \fig{fig:g_analysis}, but for the \ngn\ simulation ($f_{\NL}
  = 100$). The results from a frequentist analysis are $\hat{f}_{\NL}
  = 97, \ \sigma_{f_{\NL}}^{MC} = 20$. Using the Bayesian method, we
  obtain $\langle f_{\NL} \rangle = 99$ and $\sigma_{f_{\NL}} =
  15$. For a significant detection of \fnl, the bispectrum estimator
  shows excess variance, whereas the analysis on the basis of the
  exact Bayesian approach still provides tight error bounds.}

\twofig{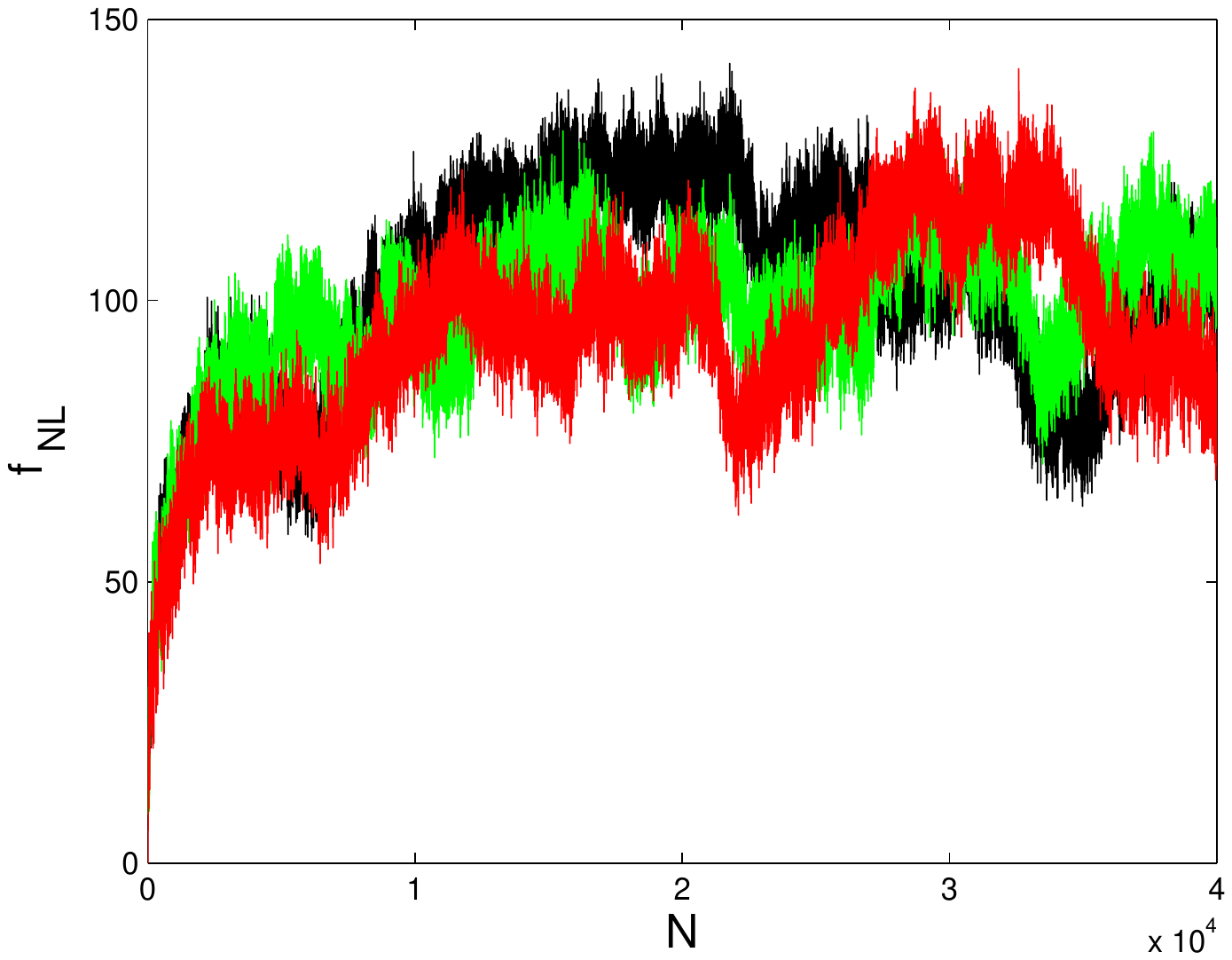}{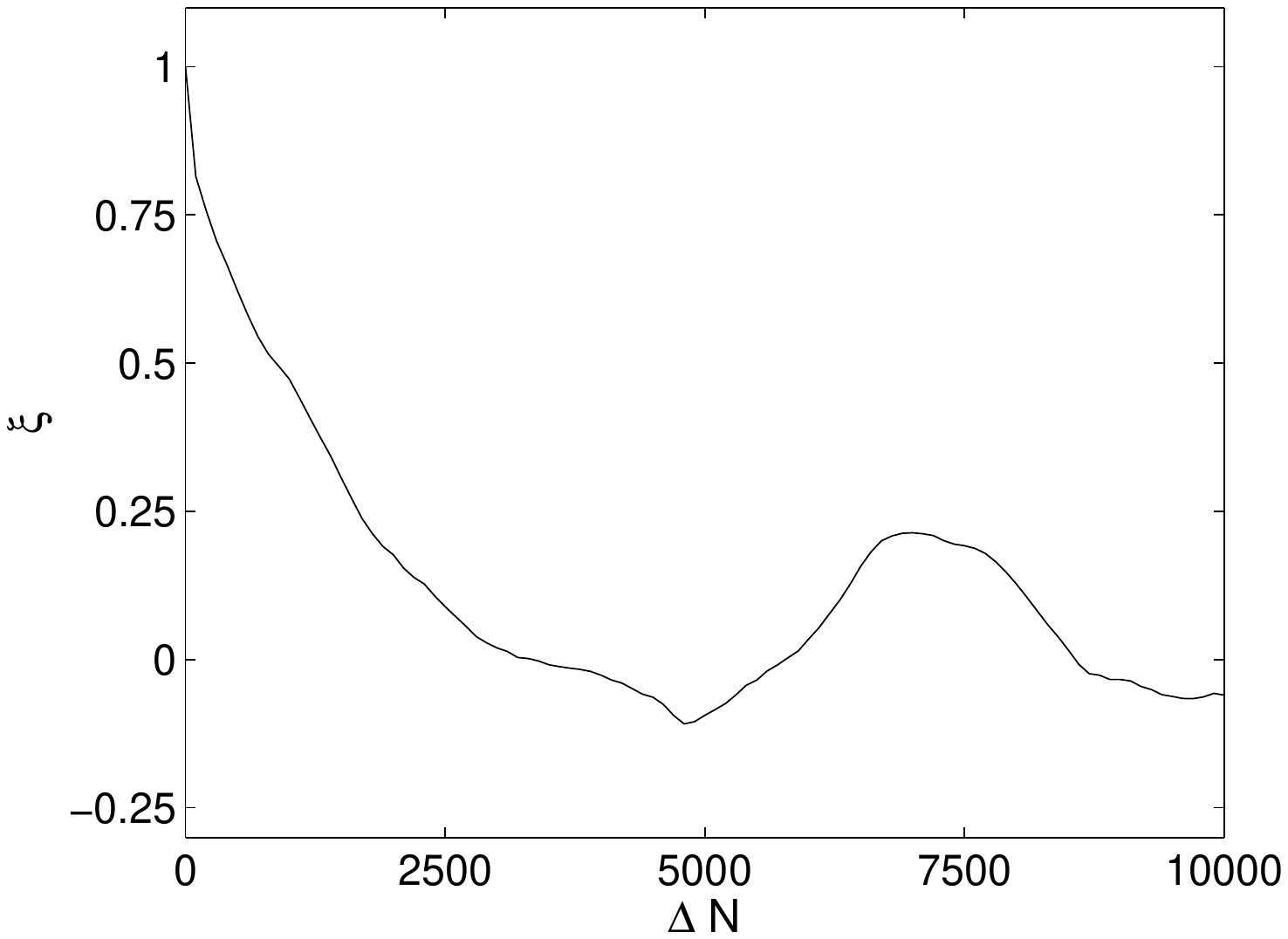}{fig:hmc_performance}%
{Performance of the sampling algorithm. \emph{Left panel:} We plot a
  random selection of three of the 32 \fnl\ sampling chains that build
  up the histogram in \fig{fig:ng_analysis}. We discarded the first
  $10\,000$ samples during burn-in. \emph{Right panel:} The
  autocorrelation function of a sampling chain.}

\section{Application to more realistic simulations}
\label{sec:wmap}

In the previous section, we have demonstrated the Bayesian approach
under idealized conditions such as isotropic noise properties and a
full sky analysis. However, applying the method to a realistic CMB
data set requires the ability to deal with spatially varying noise
properties and partial sky coverage.

In this context, a general problem is the mixture of preferred basis
representations. Whereas the covariance matrix of the primordial
perturbations can naturally be expressed in spherical harmonic space,
the noise covariance matrix and the sky mask are defined best in pixel
space. For the frequentist estimator, this is known to be problematic
as e.g.\ in the calculation of the auxiliary map $B(r, \hat{n})$ in
\eq{eq:B_map} (the Wiener filtered primordial fluctuations, see also
\eqs{eq:wf} for an equivalent, but more didactic expression), the
inversion of a combination of the two covariance matrices has to be
computed. For anisotropic noise, this can only be done by means of
iterative solvers, whose numerical efficiencies depend crucially on
the ability to identify powerful preconditioners\footnote{This can be
  very difficult, see, e.g., the discussion in
  \citet{2007PhRvD..76d3510S}}.

For the Bayesian analysis scheme as presented here, however, the
relevant equations do not contain any terms of this
structure. Therefore, the computations remain straightforward even in
the presence of arbitrary anisotropic noise properties and sky
cuts. To demonstrate this ability, we performed a reanalysis of the
simulated \ngn\ temperature map of \sect{sec:comparison}, now
superimposed by anisotropic noise as typically expected for a high
frequency WMAP channel. With these parameters, the average noise power
spectrum roughly remains at a level of about
$\mathcal{C}^{noise}_{\ell} \approx 10^{-7} \mathrm{mK}^2$, but the
noise is no longer spatially invariant. Including the KQ75y7 extended
temperature mask, we show the diagonal elements of the inverse
noise covariance matrix in \fig{fig:noise_cov}.

Again, for the analysis, we generated 32 independent Monte Carlo
chains with $140 \, 000$ samples. After discarding the first $15 \,
000$ elements during burn-in, we applied the Gelman-Rubin convergence
diagnostics to the chains and obtain a value of $R = 1.5$. The
computed mean of $\langle f_{\NL} \rangle = 90$ and the 1-$\sigma$
error of $\sigma_{f_{\NL}} = 17$ are in agreement with the input
values of the simulation. We show the constructed histogram on the
right hand panel of \fig{fig:noise_cov}, demonstrating the
applicability of the algorithm to realistic data sets.

\twofig{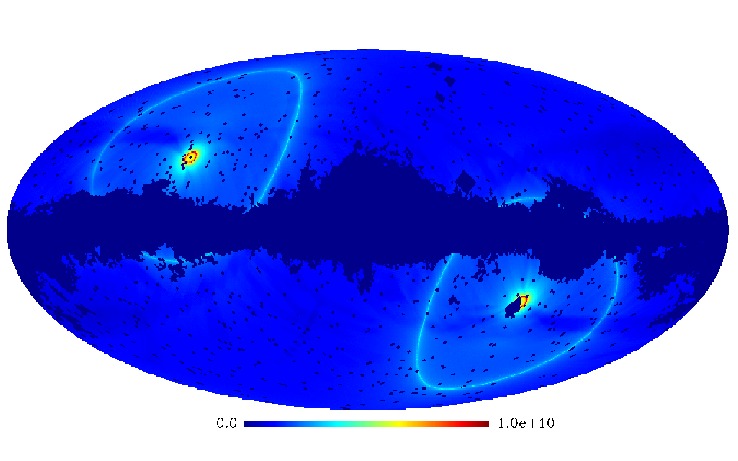}{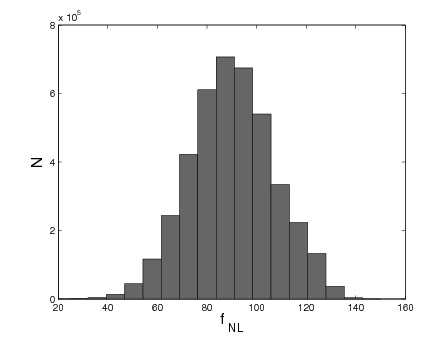}{fig:noise_cov}%
{Analysis in case of a realistic CMB experiment. \emph{Left panel:} We
  show the diagonal elements of the inverse noise covariance matrix in
  dimensionless units adopted for the more realistic simulation. When
  expressed in real space basis, off-diagonal terms vanish
  exactly. Pixel within the KQ75y7 mask are set to zero, corresponding
  to assigning infinite variance to them. \emph{Right panel:} The
  constructed posterior distribution $P(f_{\NL} | d, \theta)$ of the
  simulated map. Obtaining $\langle f_{\NL} \rangle = 90$ and
  $\sigma_{f_{\NL}} = 17$ for the mean and standard deviation,
  respectively, the input value ($f_{\NL} = 100$) gets consistently
  recovered.}

\section{Summary}
\label{sec:summary}

In this paper, we introduced an exact Bayesian approach to infer the
level of \ngy\ of local type, \fnl, from realistic CMB temperature
maps. We derived conditional probabilities for the primordial
perturbations given the data, $P(\Phi_{\LL} | d, \theta)$, and for
\fnl\ given the data and the perturbations, $P(f_{\NL} | d,
\Phi_{\LL}, \theta)$. We used Hamiltonian Monte Carlo sampling to draw
valid realizations of $\Phi_{\LL}$ from which we in turn sample
\fnl. After convergence these are samples from the full Bayesian
posterior density of \fnl\ given the data.

For a direct comparison of the newly developed scheme to the
conventional fast (bispectrum) estimator, we used simulated \gn\ and
\ngn\ CMB maps superimposed by isotropic noise. Estimates of the error
bars within the frequentist approach were derived from Monte Carlo
simulations. As a result, we find consistent outcomes between the two
approaches for the analyzed \gn\ map, in agreement with the fact that
the fast estimator is optimal in the limit of vanishing \ngy. In the
\ngn\ case, however, the advantage of the exact Bayesian approach
becomes important. Here, the uncertainty in \fnl\ remains at the same
level as for the \gn\ simulation, whereas the frequentist technique
suffers from excess variance. Our results give the first example of an
estimator (the ``mean posterior estimator'') that saturates the
Cramer-Rao bound for \fnl\ even if the signal is detectably
non-Gaussian.

Finally, we demonstrate the applicability of the newly developed
method to a realistic data set with spatially varying noise properties
and partial sky coverage. Considering a WMAP-like noise covariance
matrix and imposing the KQ75y7 extended temperature analysis mask, we
analyze a \ngn\ simulation and recover the input value consistently.

In the limit of undetectable non-Gaussianity, the Bayesian approach
ought to yield the same information as the optimal bispectrum
estimator \citep{2005PhRvD..72d3003B, 2007JCAP...03..019C}. Even in
that limit it is useful as a cross-check since it is implemented in a
completely different way. Although being computationally expensive, we
conclude that the method presented here is a viable tool to exactly
infer the level of \ngy\ of local type from CMB radiation experiments
within a Bayesian framework.

\acknowledgments
We thank the anonymous referee for the comments which helped to
improve the presentation of our results. Some of the results in this
paper have been derived using the HEALPix \citep{2005ApJ...622..759G}
package. This research was supported in part by the National Science
Foundation through TeraGrid resources provided by NCSA under grant
number TG-MCA04N015.  BDW is partially supported by NSF grants AST
0507676 and AST 07-08849. BDW gratefully acknowledges the Alexander
v. Humboldt Foundation's Friedrich Wilhelm Bessel Award which funded
part of this work.

\bibliographystyle{aa}
\bibliography{literature}

\end{document}